\documentclass[aps,pra,reprint,nofootinbib]{revtex4-1}
\usepackage{blindtext}
\usepackage[english]{babel}
\usepackage{graphicx}
\usepackage{float}
\usepackage{xcolor}

\usepackage{amsmath} 
\usepackage{amssymb}
\usepackage{amsfonts}

\newcommand\numberthis{\addtocounter{equation}{1}\tag{\theequation}} 

\usepackage{pgfplots}
\usepackage{tikz}

\usepackage{xspace}
\newcommand{\ket}[1]{\ensuremath{|#1\rangle}\xspace}
\newcommand{\bra}[1]{\ensuremath{\langle #1|}\xspace}

\DeclareMathOperator{\e}{e} 

\usepackage{mhchem}

\usepackage{tikz}
\usetikzlibrary{matrix}

\begin{document}
\title{NOON states with ultracold bosonic atoms via resonance- and chaos-assisted tunneling}
\author{G. Vanhaele}
\author{P. Schlagheck}
\affiliation{CESAM Research Unit, University of Liege, 4000 Li\`ege, Belgium}
\date{\today}

\begin{abstract}
We theoretically investigate the generation of microscopic atomic NOON states, corresponding to the coherent $\ket{N,0} + \ket{0,N}$ superposition with $N\sim 5$ particles, via collective tunneling of interacting ultracold bosonic atoms within a symmetric double-well potential in the self-trapping regime. We show that a periodic driving of the double well with suitably tuned amplitude and frequency parameters allows one to substantially boost this tunneling process without altering its collective character. The time scale to generate the NOON superposition, which corresponds to half the tunneling time and would be prohibitively large in the undriven double well for the considered atomic populations, can thereby be drastically reduced, which renders the realization of NOON states through this protocol experimentally feasible. Resonance- and chaos-assisted tunneling are identified as key mechanisms in this context. A quantitative semiclassical evaluation of their impact onto the collective tunneling process allows one to determine the optimal choice for the driving parameters in order to generate those NOON states as fast as possible.	
\end{abstract}

\maketitle
\section{Introduction}

NOON states have attracted considerable attention in the quantum physics community during the past decades. They can be seen as specific examples of Schr\"odinger cat states featuring a coherent superposition of macroscopically different quantum states \cite{Schrodinger1935}. Most generally, they are realized within bosonic two-mode systems and correspond to the superposition $e^{i \varphi_1}|N,0\rangle + e^{i \varphi_2}|0,N\rangle$ of the two quantum states $|N,0\rangle$ and $|0,N\rangle$ where one of the modes is populated with $N$ quanta while the other one is empty.
While being an important resource for quantum information processing, the giant entanglement that is inherent in those NOON states renders their experimental realization notoriously difficult. Impressively large NOON states with $N \leq 10$ quanta were nevertheless created with qubits in superconducting circuits \cite{Song2018}, with photons \cite{Afek2010} as well as with phonons in ion traps \cite{Zhang2018}, to mention a few examples.

The realization of NOON states with ultracold bosonic atoms is of particular interest as it enables the creation of Schr\"odinger cat states made of massive matter. A key feature of atoms is that they generally interact with each other. Rather than being a nuisance, the presence of atom-atom interaction can be seen as an important asset since it allows one to conceive protocols for the generation of the NOON superposition that are based on the internal dynamics of the atomic gas. Quite straightforwardly, for instance, the ground state of an attractively interacting gas of bosonic atoms within a perfectly symmetric double well potential is a NOON state where all atoms are located in the same (left or right) well. If attractive interactions are to be avoided due to the enhanced instability of the atomic gas, one can resort to a two-component Bose gas with an important inter-component interaction strength, which also features a NOON state as the ground state \cite{Cirac1998}. However, preparing such ground states through cooling techniques is a rather challenging task since extremely low temperatures are required in order to distill the symmetric NOON superposition $|N,0\rangle + |0,N\rangle$ with respect to its nearly degenerate antisymmetric counterpart $|N,0\rangle - |0,N\rangle$.
As an alternative to ground-state cooling, a number of dynamical schemes have been proposed in order to create a NOON state with ultracold bosonic atoms. A widely discussed scenario consists in inducing a suitable phase shift \cite{Sorensen2001} between the two wells of a double well potential hosting a Bose-Einstein condensate of repulsively interacting atoms. The subsequent dynamical redistribution of atoms between the wells will, after a specific evolution time, give rise to NOON superpositions with a fairly decent purity \cite{Gordon1999,Dunningham2001,Micheli2003,Mahmud2003, Mahmud2005, Zibold2010}. Other interesting proposals are based on well-designed measurement schemes to distill the NOON superposition \cite{Ruostekoski1998, Mazets2008,Cable2011}, on generating the NOON superposition via an adiabatic passage through an excited-state quantum phase transition \cite{Bychek2018}, on two-component Bose-Einstein condensates in double well potentials \cite{Teichmann2007} or on splitting processes of solitonic wave packets transporting attractively interacting atoms \cite{Weiss2009,Streltsov2009}.

A conceptually simple and hence rather appealing protocol, which was also discussed in Ref.~\cite{Carr2010}, consists in preparing a Bose-Einstein condensate in one of the two wells of a perfectly symmetric double well potential and then waiting until it starts tunneling to the other well. This protocol would have to be implemented in the so-called self-trapping regime \cite{Smerzi1997,Milburn1997,Steel1998,Spekkens1999,Ostrovskaya2000,Leggett2001, Albiez2005, Ananikian2006,Fu2006,Julia-Diaz2010} where the (repulsive) atom-atom interaction is so strong as compared to the inter-well hopping that the atoms are inhibited to go to the other well individually, due to a mismatch of their chemical potentials in the two wells, and can only tunnel there in a collective manner, by means of a quantum sloshing process as it was termed in Ref.~\cite{Carr2010}. A nearly perfect NOON state is then realized at half the time that it takes to undergo a complete tunneling process of the condensate to the other well. The main drawback of this proposal, besides the requirement of perfect symmetry, is that the evolution time scale needed to reach this NOON state is extremely long (see, e.g., \cite{Salgueiro2007}), such that it would most generally exceed the typical decay time of the condensate due to three-body collisions.

In this paper, we show how to substantially speed up this quantum sloshing process and hence also the production of NOON states by subjecting the double well potential to a periodic shaking \cite{Watanabe2010,Watanabe2012}. The amplitude and frequency of this shaking are to be chosen such that a significant layer of chaotic motion is induced within the classical phase space of the mean-field dynamics within this two-mode system, without appreciably altering the classical orbits that support the $|N,0\rangle$ and $|0,N\rangle$ states. The transition matrix element between these two quantum states is then drastically enhanced owing to chaos-assisted tunneling \cite{Lin1990,Bohigas1993,BohigasTom1993,Tomsovic1994,Leyvraz1996} which proceeds via a perturbative coupling of the ``regular'' $|N,0\rangle$ and $|0,N\rangle$ states to the manifold of ``chaotic'' states that are strongly connected with each other due to the presence of the periodic driving. This regular-to-chaotic coupling can be further boosted by the presence of one or several important nonlinear resonances in the classical dynamics \cite{Brodier2001,Brodier2002}, thereby giving rise to a combination of resonance- and chaos-assisted tunneling \cite{Eltschka2005,Schlagheck2006,Mouchet2006,Lock2010,Schlagheck2011}. As we demonstrate below, NOON states can thereby be generated at drastically reduced evolution time scales, and this without notably affecting their purity.

Our paper is organized as follows. Section II is devoted to the description of the physical system under consideration, namely ultracold bosonic atoms trapped in a double-well potential. The notions of NOON state and NOON time are introduced. In Section III, the presence of a periodic shaking of this double potential is considered. We show that this periodic shaking gives rise to an enhancement of tunneling due to chaos- and resonance-assisted tunneling. In Section IV we show that it is also possible to drastically decrease the NOON time in the near integrable regime, where chaos is not present.

\section{NOON states in the two-mode Bose-Hubbard model}
The physical framework consists in ultracold bosonic atoms contained in a double well potential. The latter can be experimentally realized through an optical lattice restrained to two sites by means of an additional harmonic confinement  \cite{Albiez2005} or through superlattice techniques \cite{Folling2007}. These two sites are coupled with each other through $J$, the hopping parameter. The on-site two-body interaction $U$ tends to localize atoms on a specific site. If the temperature is sufficiently low, the system stays on the lowest energy band, featuring one orbital per lattice site, and the Hamiltonian reads \cite{Milburn1997,Leggett2001,Dalton2012}
\begin{align}\label{Hubbard}
\hat H_0=&-J (\hat a_1^\dagger \hat a_2+\hat a_2^\dagger \hat a_1)+\frac{U}{2}(\hat a_1^\dagger \hat a_1^\dagger \hat a_1 \hat a_1+\hat a_2^\dagger \hat a_2^\dagger \hat a_2 \hat a_2 ).
\end{align}

\begin{figure}[p]
	\includegraphics[width=0.5\textwidth]{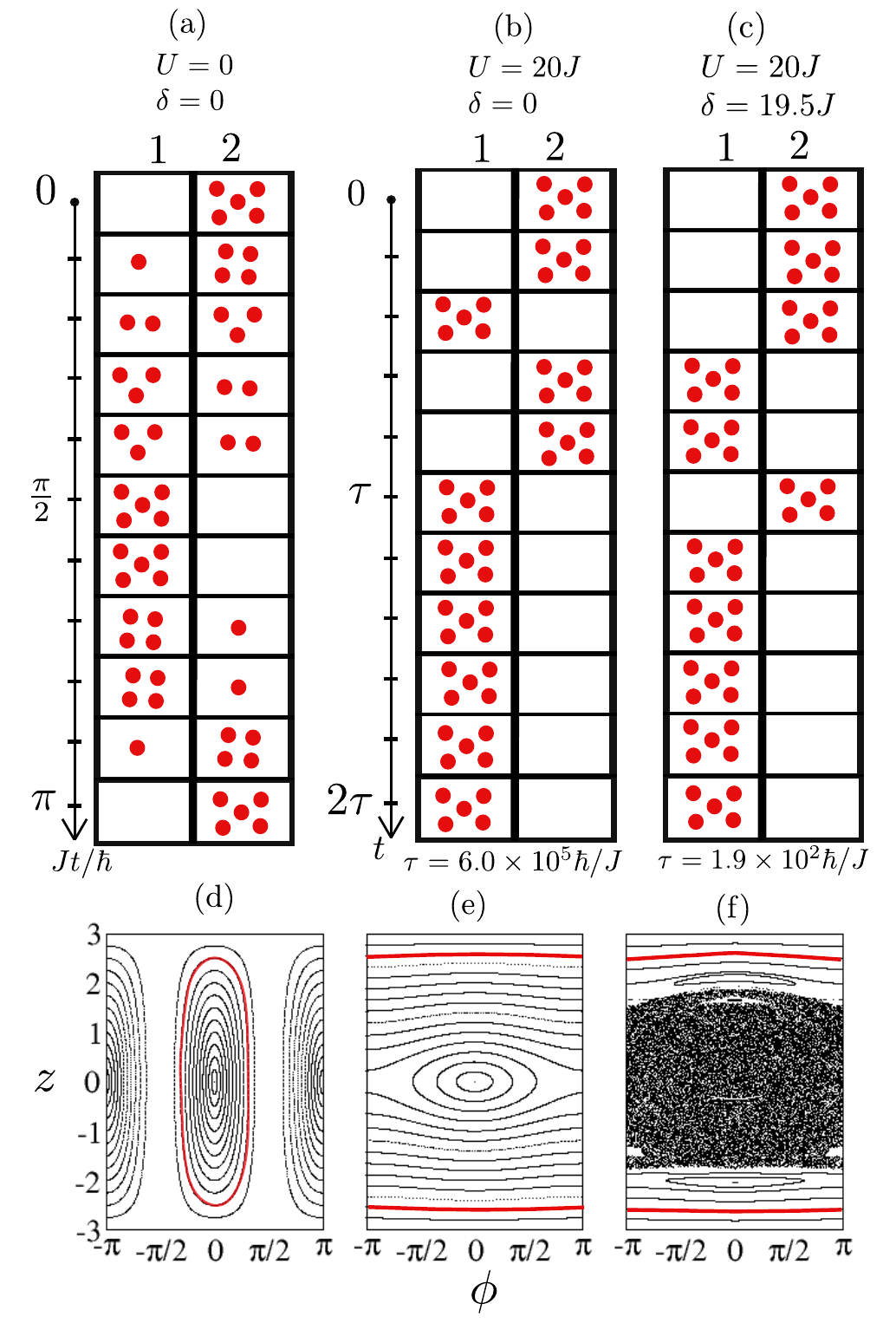}
	\caption{ Sequential (a) and collective tunneling (b and c) of ultracold bosonic atoms in a double-well potential, in the absence (a and b) and in the presence (c) of a periodic driving. In the upper panels (a-c), a measurement process of the number of atoms within individual wells is numerically simulated for various evolved times, based on the numerical time evolution generated by the Hamiltonian (\ref{quantum_H_t}).
		The lower panels (d-f) represent the corresponding phase spaces in the mean field approximation (see Eqs. (\ref{discrete_GP_1}) and (\ref{discrete_GP_2}) or the Hamiltonian (\ref{mean-field})). The population imbalance and the phase difference between site 1 and 2 are respectively $z=(N_1-N_2)/2$ and $\phi=\theta_1-\theta_2$. The red trajectories are the classical counterparts of the upper quantum dynamics. (a,d) In the absence of interaction, tunneling occurs one by one, leading to Josephson oscillations. (b,e) In the presence of interaction ($U=20J$), collective tunneling takes place, but on a very long time scale, $\tau=6.0\times 10^5\hbar/J$. The phase space displays two qualitatively different dynamics, namely Josephson oscillation for low population imbalances and self-trapping for high population imbalance. (c,f) In the presence of a periodic shaking characterized by an amplitude $\delta/J=19.5$ and a frequency $\hbar\omega/J=20$, collective tunneling occurs on a much shorter time scale, $\tau=1.9\times 10^2\hbar/J$. The presence of a central chaotic layer and a 1:4 resonance roughly situated at $z=\pm2$ is able to explain the decrease of the NOON time.
	}
	\label{experimental_simulation}
\end{figure}

 The panel (a) of the figure \ref{experimental_simulation} displays a numerical simulation of the in-situ measurement of the atoms in the left and right well, as a function of evolution time. This calculation was done with the initial condition $\ket{0,5}$ i.e. with 5 particles, all of them on site 2. For a non-interacting gas ($U/J=0$), this initial state gives rise to individual atomic transfer in a periodic way (panel (a) of the figure \ref{experimental_simulation}) known as Josephson oscillations \cite{Zapata1998,Levy2007}. In this case, the NOON state does not occur.

For a sufficiently high ratio $U/J$, all five atoms initially stay localized on site 2 in the so called self-trapping regime. Indeed, the mismatch between the chemical potentials of the two sites inhibits individual transfer of atoms from the state $\ket{0,5}$ to the other states (see panel (b) of the figure \ref{experimental_simulation}). On a very long time scale, giant transfer of matter through collective tunneling will nevertheless take place due to the quasidegeneracy between the symmetric and the antisymmetric superposition of $\ket{0,N_p}$ and $\ket{N_p,0}$ (with $N_p=5$, the total number of particles), whose eigenvalues are respectively $\epsilon^ +$ and $\epsilon^-$. This process occurs through a NOON state, meaning that the wavefunction can be written as
\begin{equation} \label{pure_NOON}
\ket{\text{NOON}}=\cos\left(\frac{\Delta \epsilon} {2\hbar} t\right) \ket{0,N_p}+i \sin\left(\frac{\Delta\epsilon} {2\hbar} t\right)\ket{N_p,0}
\end{equation}
with $\Delta\epsilon=\epsilon^ - -\epsilon^ +$, the level splitting. The expression (\ref{pure_NOON}) is an approximation that neglects the perturbative admixture of the other states $\ket{1,4}$, $\ket{2,3}$, $\ket{3,2}$ and $\ket{4,1}$. The time needed to obtain a perfectly balance superposition between the two quasimodes $\ket{0,5}$ and $\ket{5,0}$ reads 
\begin{equation}\label{tau}
\tau=\dfrac{\pi\hbar}{2|\Delta \epsilon|}.
\end{equation}

For $N_p=5$, $U/J=20$ and $\delta/J=0$ (panel (b) of the figure \ref{experimental_simulation}), the numerical result gives $\tau= 6.0 \times 10^{5} \hbar/J$. This NOON time can be expressed in physical units for the atomic species $\ce{^{87}Rb}$ characterised by a mass $m=1.443\times10^ {-25}$kg and a s-wave scattering length $a_s=5.313$nm. If the optical lattice is produced by a laser with the wavelength  $\lambda=1064$nm, we obtain $\hbar/J=4.4\times 10^{-3}$s and the NOON time becomes $\tau=2.6\times 10^3$s (see Appendix \ref{an_parameter_evaluation} for the derivation). This time is prohibitively large in comparison with the typical lifetime of a  condensate in an optical lattice equivalent to roughly 10s \cite{Andersen2008}. The situation gets worse for an even larger number of particles, and this in a dramatic manner: with the same lattice parameters, we obtain $\tau=2.2\times10^5$s for $N_p=6$, and $\tau=2.3\times 10^7$s for $N_p=7$. This is not surprising as the NOON time is expected to increase exponentially with the number of particles.

A semiclassical analysis can help to understand this phenomenon. The individual transfer of particles between sites in the Josephson regime is a classical transition successfully described by the Gross-Pitaevskii equation. The latter fails to reproduce the collective transfer of particles. The lower panels of figure \ref{experimental_simulation} show the phase spaces related to the quantum numerical simulations in the corresponding upper panels. The red curves represent the classical tori on which the symmetric and antisymmetric combination of $\ket{0,5}$ and $\ket{5,0}$ are anchored. In the mean field approximation, we can extract the discrete Gross-Pitaevskii equation from the Hamiltonian
 (\ref{Hubbard}) \cite{Smerzi1997,Ananikian2006},
\begin{eqnarray}\label{discrete_GP_1}
i\hbar \dfrac{\partial \psi_1}{\partial t}&=-J\psi_2+U|\psi_1|^2 \psi_1
\\
i\hbar \dfrac{\partial \psi_2}{\partial t}&=-J\psi_1+U|\psi_2|^2 \psi_2. 
\label{discrete_GP_2}
\end{eqnarray}
In this framework, a condensate amplitude can be associated to the site 1 and 2, $\psi_{1,2}=\sqrt{N_{1,2}+1/2} \e^ {i\theta_{1,2}}$ where $N_j$ and $\theta_j$ are respectively the number of particles and the phase of the site $j$. The transformation to $(z,\phi)$ coordinates following $z=(N_1-N_2)/2$ and $\phi=\theta_1-\theta_2$ enables to introduce the Hamiltonian related to the equations (\ref{discrete_GP_1}) and (\ref{discrete_GP_2}),
\begin{equation}
  H_0(z,\phi)=U z^2 - 2J \sqrt{\left(N_Q/2\right)^2 -z^2 } \ \cos\phi,
\end{equation}
 with $N_Q=N_p+1$ \cite{Smerzi1997}. 
 
 For a non-interacting gases all initial conditions give rise to Josephson oscillations (panel (a) of figure \ref{experimental_simulation}) while, for finite $U/J\neq 0$, self-trapping and Josephson oscillation coexist in the phase space \cite{Smerzi1997, Albiez2005} (panel (e) of the figure \ref{experimental_simulation}). The mean-field approximation (\ref{discrete_GP_1}) and (\ref{discrete_GP_2}) is able to characterize those two classical regimes, but fails to reproduce the collective tunneling that takes place in the self-trapping regime in the panel (b) of the figure \ref{experimental_simulation}. For this a quantum or a semiclassical approach is needed to model the giant transfer of matter that takes place through tunneling across the dynamical barrier. This gives rise to an exponential scale of the NOON time with $N_p$, which explains why this time is so large.
 


\section{Chaos- and resonance-assisted tunneling}
The NOON time can be significantly reduced by means of a periodic driving. In this case, the system is no longer integrable meaning that chaos and nonlinear resonances can emerge from the phase space. As an abundant literature suggests it \cite{Brodier2001,Brodier2002,Eltschka2005,Schlagheck2006,Mouchet2006,Lock2010,Schlagheck2011,Mertig2016,Fritzsch2017}, these two structures and their positions in the phase space have a huge impact on the dynamical tunneling, whose inherent time can be reduced by several orders of magnitude.

A specific driving which effectively corresponds to a periodic tilting of the double well potential is introduced through
\begin{equation}\label{quantum_H_t}
\hat H (t) = \hat H_0+ \delta \cos(\omega t) (\hat a_1^\dagger \hat a_1 -\hat a_2^\dagger \hat a_2).
\end{equation}
In a time periodic system, the Floquet theory \cite{Eckardt2005} is a suitable framework to compute the quasienergies, $\epsilon^\pm$, in order to determine the NOON time (see (\ref{tau})). Any state $\ket{\phi(t)}$ can be decomposed in the time-periodic basis $\{\ket{u_\nu^ \sigma(t)} = \ket{u_\nu^\sigma(t+2\pi/\omega)}\}$,
\begin{equation}
\ket{\phi(t)} = \sum_{\sigma=\pm}\sum_{\nu=1}^{D^ \sigma} c_\nu^ \sigma (t_0) \e ^{-\frac i \hbar \epsilon_\nu^ \sigma (t-t_0)} \ket{u_\nu^ \sigma (t)},
\end{equation}
with $c_\nu^ \sigma (t_0) = \langle u_\nu^\sigma (t_0) \ket{\phi(t_0)}$. $D^\pm$ are respectively the dimension of the symmetric and antisymmetric blocks such that $D^+ + D^ -$ is the Hilbert space dimension of the unperturbed system. The eigenvalue Schr\"odinger equation reads \cite{Eckardt2005}
\begin{equation}\label{FloquetEquation}
\left(\hat H(t)-i\hbar\partial_t\right)\ket{u^\sigma_\nu (t)}=\epsilon^\sigma_\nu\ket{u^\sigma_\nu (t)}.
\end{equation}
The solution of the Fourier series of this equation leads to the quasienergies $\epsilon^\sigma_\nu$ and the Floquet eigenstates, in terms of the Fourier coefficients of $\ket{u^\sigma_\nu (t)}$.

The panels (c,f) of the figure \ref{experimental_simulation} represent the same lattice parameters as in panels (b,e) of the same figure except that this system is periodically perturbed ($\delta/J=19.5$ and $\hbar\omega/J=20$). The upper and lower margins of these phase space plots are very similar as they both display a self-trapping structure, while the center of (f) is dominated by a large chaotic sea.
This prominent chaotic sea, which is characterized by an admixture between the quasimodes $\ket{1,4}$, $\ket{2,3}$, $\ket{3,2}$ and $\ket{4,1}$ facilitates the transition from $\ket{0,5}$ to $\ket{5,0}$. Moreover, a 1:4 resonance in $z_{1:4}\simeq\pm 2$ is symmetrically located between the quasimodes $\ket{0,5}$ and $\ket{1,4}$, making easier a transition to the chaotic sea. These two new structures in the phase have an impact on the splitting that determines the NOON time $\tau = 1.9\times 10^2\hbar/J$. For $\ce{^{87}Rb}$ and the experimental parameters mentioned in the previous section, this NOON time amounts to $\tau=0.84$s. The qualitative dynamics is the same as in the unperturbed system (panel (b) of the figure \ref{experimental_simulation}), but the external shaking is able to reduce the NOON time by more than three orders of magnitude. This time can be compared to the period $T=2\pi/\omega=1.4 \times 10^{-3}$s of the driving, meaning that the system must carry out roughly 600 cycles to reach the NOON state.

An analysis of the phase space gives valuable information for the optimal choice of the external frequency leading to a significant increase of tunneling. A quantitative insight into the role of $\omega$ can be obtained from the mean-field Hamiltonian related to the phase space displayed in the panel (f) of the figure \ref{experimental_simulation},
\begin{equation}\label{mean-field}
H(z,\phi, t)=H_0(z,\phi) + 2\delta   \cos\left(\omega t\right)  z.
\end{equation}
The following strategy is adopted to determine the choice of the external frequency. Firstly, a range of $\omega$ can be chosen in order to create a central chaotic layer (see the panel (f) in comparison with the panel (e), both in the figure \ref{experimental_simulation}). Then, a more refined analysis is needed in order to act on the tunneling of a specific pair of quasimodes. Owing to the periodicity of the perturbation in the Hamiltonian (\ref{mean-field}), the phase space can be represented through a stroboscopic section. If this perturbation is not too strong, the Kolmogorov-Arnold-Moser theorem \cite{Tabor1989,Lichtenberg2013,Wimberger2014} states that tori with incommensurable winding numbers $\alpha=\omega/\Omega$, with $\omega$ the frequency of the driving and $\Omega$ the frequency of the torus, will be preserved. Conversely, the Poincar\'e-Birkhoff theorem states that the tori with commensurable $\alpha=r/s$, with integer $r$, $s$, will be destroyed and will lead to nonlinear resonances with $r$ pairs of stable and unstable fixed points and chaos close to the unstable fixed points \cite{Tabor1989,Lichtenberg2013,Wimberger2014}. 

The location of a nonlinear resonance can be precisely determined by means of classical perturbation theory. In the phase spaces of the figure \ref{experimental_simulation}, the torus $z(\phi)$ is characterized by the action variable $I=1/(2\pi) \int_{0}^{2\pi} z(\phi) d\phi$. For $\delta=0$, the canonical perturbation theory \cite{Lichtenberg2013} at the order 1 (valid for $N_Q U /J\gg1$ and for high population imbalances) gives rise to
\begin{equation}
   z(\phi)\simeq I +\dfrac J{(U I)} \sqrt{\left( N_Q /2\right)^2-I^2} \cos\phi
\end{equation}
  and the new Hamiltonian 
  \begin{equation}
  H_0(I)\simeq U  I^2.
  \end{equation}
   That enables to compute analytically the frequency of the tori via the relation $\hbar\Omega =  dH_0(I)/dI  \simeq  2U I$. The destroyed tori display a commensurable winding number $\alpha={\omega}/{\Omega_{r:s}}=r/s$ where $\Omega_{r:s}\simeq 2U  I_{r:s}/\hbar$ is the frequency of the destroyed torus. The external frequency that must be applied in order to build an $r$:$s$ resonance characterised by an action $I_{r:s}$ (meaning that the resonance will be roughly situated in $z=I_{r:s}$) is obtained through the relation
\begin{equation}\label{external_frequency}
\hbar\omega \simeq \frac r s \ 2UI_{r:s}.
\end{equation}
This relation is sufficient to find a suitable resonant frequency that decreases the NOON time. $\hbar\omega/J=20$ produces a 1:4 resonance in $z_{1:4}\simeq2$ that is symmetrically located between $\ket{5,0}$ and $\ket{4,1}$. Moreover, the interplay between the 1:1 resonance in $z_{1:1}\simeq0.5$ and the central island gives rise to the chaotic layer as displayed in the panel (f) of figure \ref{experimental_simulation}. The Appendix \ref{sc_evaluation} presents semiclassical evaluations of the NOON time in the near-integrable regime and in the mixed regime by taking into account the presence of the chaotic sea. Note that even though chaos- and resonance- assisted tunneling is able to reproduce the order of magnitude of the NOON time, this semiclassical theory is not sufficiently precise for experimental predictions, and a pure quantum analysis becomes necessary. Indeed, the semiclassical estimate $\tau=1.4\times 10^2\hbar/J$ (see Appendix \ref{sc_evaluation}) must be compared to the exact NOON time $\tau=1.9 \times 10^2\hbar/J$ obtained with our numerical simulations. 

\begin{figure}[!h]
	\centering
	\includegraphics[width=0.5\textwidth]{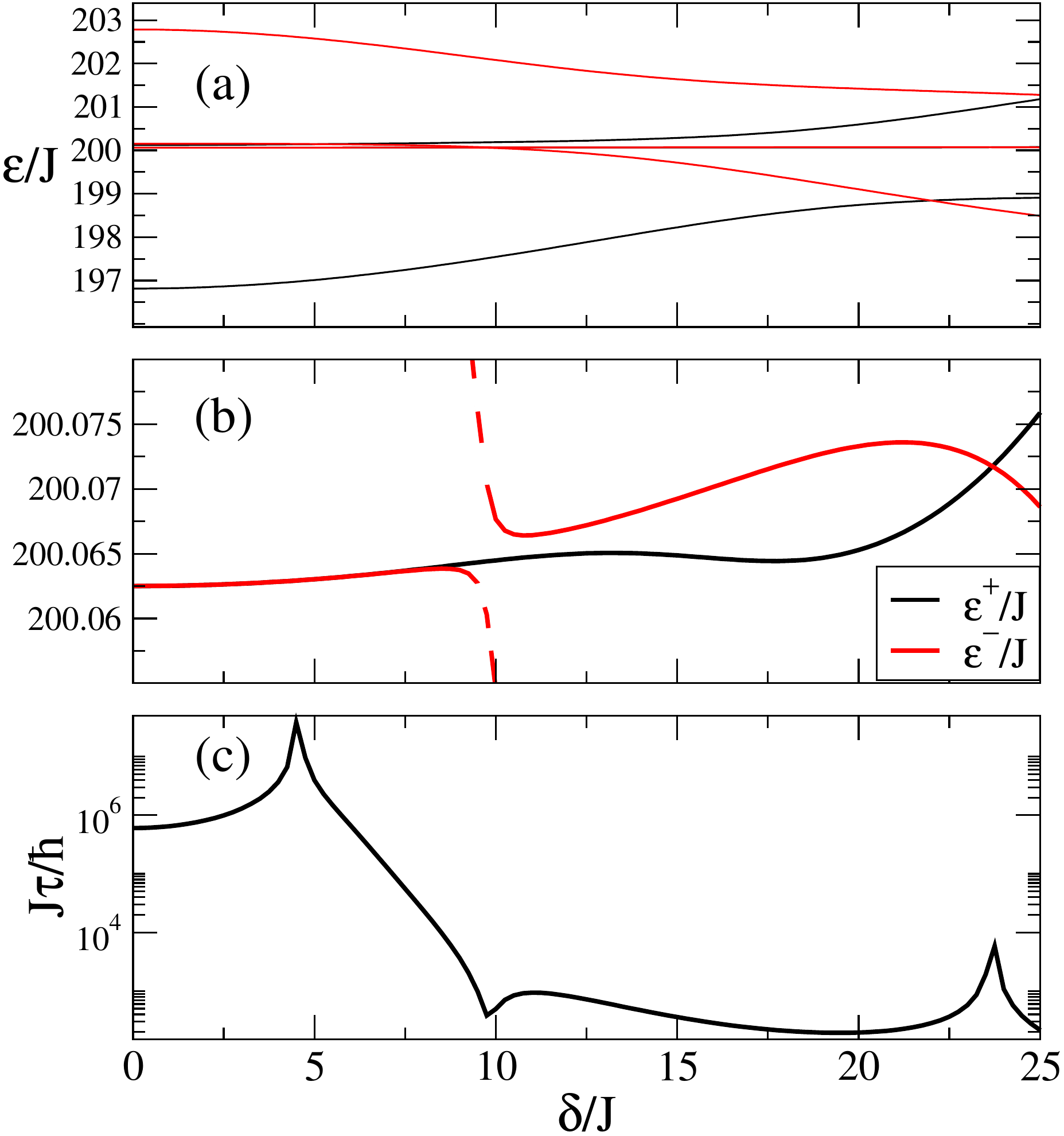}

	\caption{
		The NOON doublet in the Floquet spectrum, as a function of the driving amplitude. Panel (a) represents a block of the periodic Floquet spectrum (with even-parity levels being marked in black and odd-parity levels in red), panel (b) displays a zoom of this block and panel (c) shows the NOON time calculated from the panel (b) and the relation (\ref{tau}). The minimal $\tau$ appears roughly in $\delta/J\simeq 19.5$, which was used to produce the panels (c,f) of the figure \ref{experimental_simulation}. This result is robust in a sense that a range between $\delta/J\simeq12$ and 22 can be chosen to observe the reduction by several orders of magnitude of the NOON time. 
		Parameters: $N_p=5$, $U/J=20$ and $\hbar\omega/J=20$. }
	\label{NOON_time}
\end{figure}

In order to find an optimal value of $\delta$, all the other parameters are fixed and the NOON time is computed as a function of $\delta$. The panel (c) of the figure \ref{NOON_time} suggests that the minimum NOON time is reached for $\delta/J \approx 19.5$. This result is robust meaning that there is a range between approximately $\delta/J = 12$ and $22$ where the decrease of the NOON time is observable. This range begins roughly at a level repulsion near  $\delta/J= 10$ for which the doublet of the symmetric and antisymmetric combination of $\ket{0,5}$ and $\ket{5,0}$ is crossed by the antisymmetric combination of $\ket{1,4}$ and $\ket{4,1}$ (see the dashed line of figure \ref{NOON_time}). After that, the splitting gains several orders of magnitude. The more $\delta/J$ is increased, the more the chaotic sea is large. For $\delta/J>$ 25, the quasimodes $\ket{0,5}$ and $\ket{5,0}$ begin to be diluted in the central chaotic layer. Beyond this point, the notions of splitting and NOON time are no longer meaningful. 

	\begin{figure}[h!]
		\centering
		\includegraphics[width=0.5\textwidth]{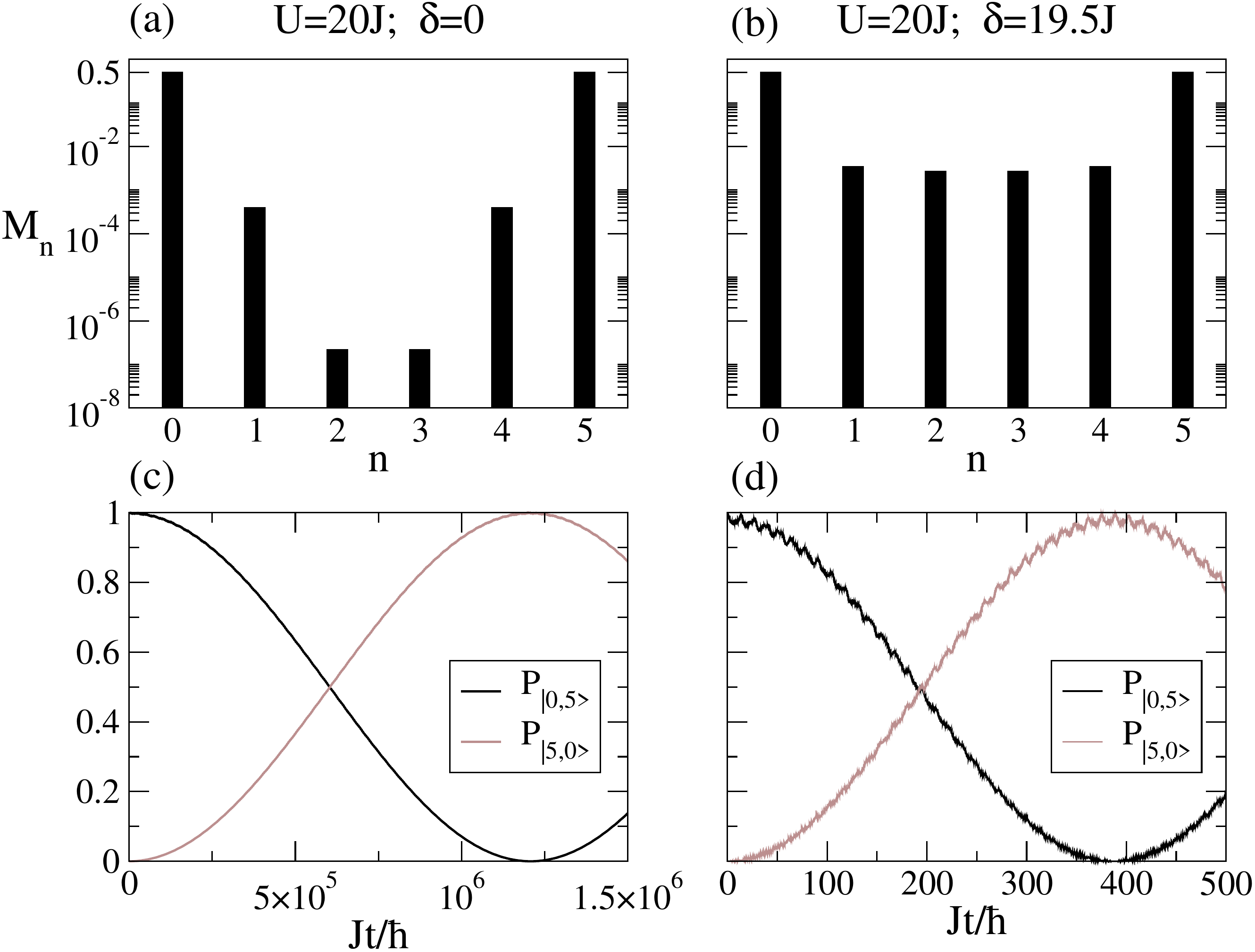}
		\caption{Overlap between the quasimodes $\ket{n,N_p-n}$ ($n=0,1,2,3,4,5$) and the Floquet eigenstate according to the relation (\ref{purity_Mnu}). Even for the perturbed system (panel (b)), the two-level approximation can still be justified (see equation (\ref{pure_NOON})) as $M_0=M_{N_p}\approx 0.5$. The lower part displays the detection probabilities knowing that the system is prepared in $\ket{0,5}$. While the left panels shows the unperturbed case, the right panels display the situation with a frequency $\hbar\omega/J=20$ that produces a chaotic layer and a 1:4 resonance (see panel (f) of the figure \ref{experimental_simulation}). 			
		}
		\label{proba_mesure_chaos}
	\end{figure}
	
In view of the latter considerations, a legitimate preoccupation could be the implication of the external driving on the purity of the NOON state. Indeed, the driving tends to decrease the NOON time, but tends also to increase the admixture between the NOON state and the other quasimodes.
To obtain a time-independent indicator of the purity of the NOON state, a simplified definition of the purity,
\begin{equation}
  p= M_0+M_{N_p},
\end{equation}
based on the overlap 
\begin{align}\label{purity_Mnu}
&M_{n} = \dfrac 1 T \sum_{\sigma=\pm} \int_{0}^{T} |\langle n,N_p-n | u_\nu^ \sigma (t) \rangle |^2 \ dt
\end{align}
is introduced. This definition of purity can formally be identified with the time average of the expectation value $\mathrm{Tr}[\hat{P}\hat{\rho}]$ of the projector to the NOON doublet with respect to the density matrix $\hat \rho = |u_\nu^+(t)\rangle\langle u_\nu^+(t) | + |u_\nu^-(t)\rangle\langle u_\nu^-(t) |$. The time average over one driving period allows one to eliminate small periodic oscillations of the purity related to micromotion, which are not of interest here. The panel (b) of the figure \ref{proba_mesure_chaos} suggests that the resulting purity is roughly equal to $p=0.99$. So the two level approximation seems to hold. Nevertheless, it is true that the introduction of the perturbation decreases the purity as this one is better than $p=0.999$ for the unperturbed case (see panel (a) of the figure \ref{proba_mesure_chaos}). The lower row of figure \ref{proba_mesure_chaos} shows that the time evolution of the transition probabilities can be described in terms of two-states dynamics in both cases.

\begin{figure}[!h]
	\includegraphics[width=0.5\textwidth]{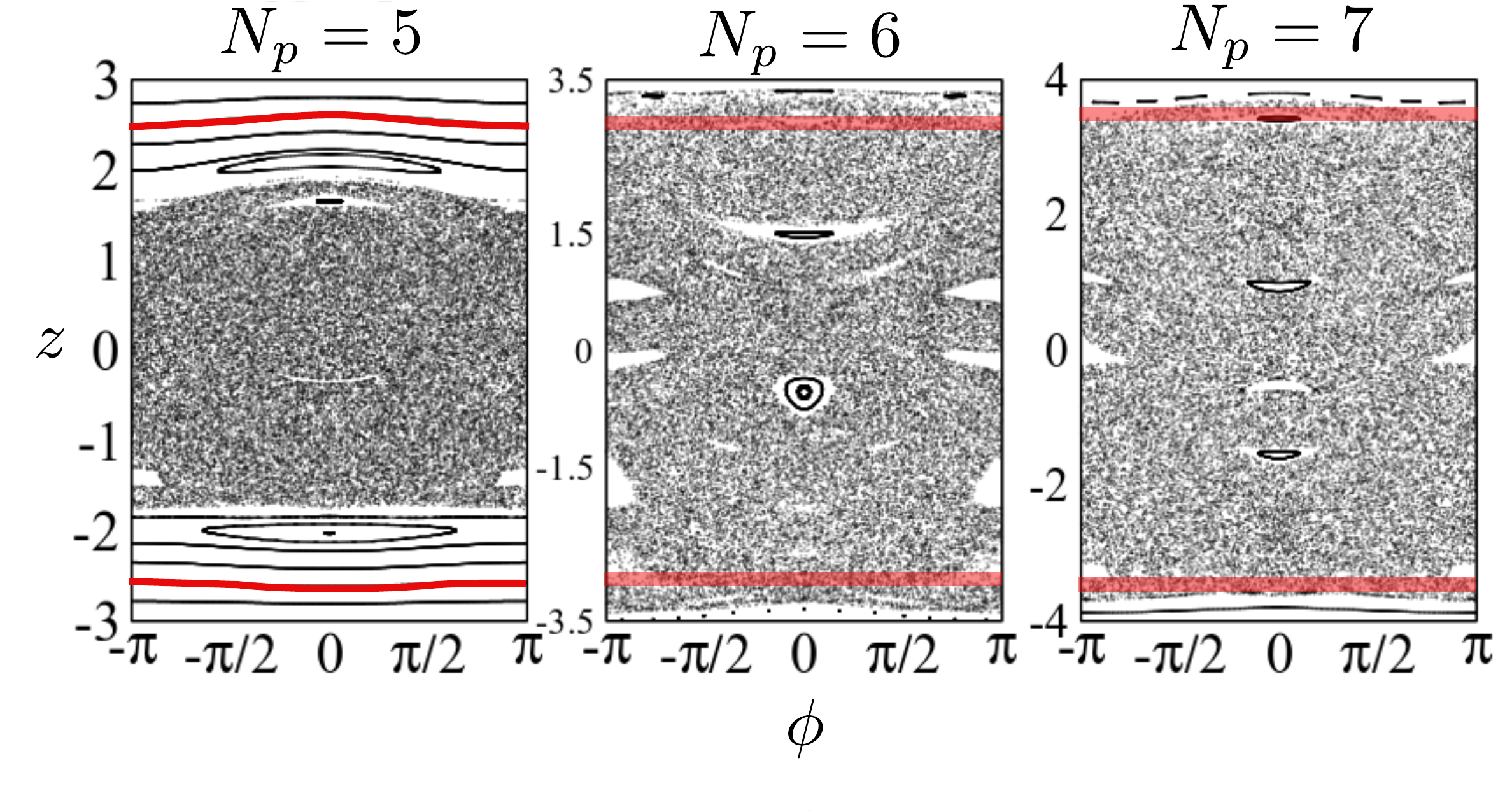}
	\caption{Evolution of the phase space with $N_p$. The nonlinear parameter $(N_p+1)U/J$ and $\hbar\omega/J$ are kept constant in order to preserve the phase space for $\delta=0$. $\delta/J$ is chosen such that a minimum of the NOON time is reached for $N_p=5$, 6, 7. We choose $\delta/J=19.5,$ 50, 44 respectively. Note that for $N_p=6$ and 7 the NOON states (highlighted by red shading on the phase space) are separated from the chaotic sea by a partial barrier.
	} \label{scaling}
\end{figure}

As pointed out in the previous section, the splitting between two related quasimodes is known to decrease exponentially with $N_p$, leading to an exponential increase of the NOON time.  This is also the case for $\delta\neq0$, but the increase is not so extreme. In figure \ref{scaling}, $N_p$ is increased while keeping the nonlinear parameter $(N_p+1)U/J$ and $\hbar\omega/J$ constant. From an experimental point of view, this particular scaling can be achieved by an adaptation of the lattice parameters without the need of modifying $a_s$. Keeping the nonlinear parameter constant enables to preserve the same phase space for $\delta=0$.

Figure \ref{scaling} represents three phase spaces for $N_p=5,$ 6, 7 with $\delta/J=19.5,$ 50, 44 respectively, which correspond to the minima of the NOON time $\tau$. For $N_p$=6 and 7, the NOON states (in red) appear to be inside the chaotic region. However, they are still isolated from the chaotic part of the Floquet spectrum by the presence of an important partial barrier in the phase space \cite{BohigasTom1993,Tomsovic1994,Schlagheck2011}. As shown in Table \ref{table_scaling}, the NOON time with a perturbation increases with $N_p$ as expected. Nevertheless, this increase is slowed down in comparison to the one in the unperturbed case. In addition to reducing the NOON time, the external perturbation is also able to decrease the slope of $\tau$ as a function of $N_p$ (see Table \ref{table_scaling}). Resonance- and chaos-assisted tunneling therefore opens interesting perspectives to create increasingly big entangled states.

\begin{table}[!h]	
		$\begin{array}{c|ccc}
			N_p & J\tau_{\delta=0}/\hbar \quad  & J\tau_{\delta\neq0}/\hbar \quad   & p_{\delta\neq0}\\
			\hline
			5 & 6.0\times 10^5 & 1.9\times 10^2 & 0.988 \\
			6 & 2.3\times 10^7 & 5.5\times 10^2 & 0.968\\
			7 & 9.2\times 10^8 & 1.4\times 10^3 & 0.987
		\end{array}$
		\caption{Comparison of the behaviours of the NOON time as a function of the total number of particles $N_p$ in the unperturbed and perturbed systems. The NOON time is expected to exponentially increase with the semiclassical parameter $N_p$. The third column shows that this increase is softened by the external perturbation, suitably tuned in order to reach a minimum of the NOON time. These minima are obtained at $\delta/J=19.5,$ 50, 44 for $N_p=5,$ 6, 7 respectively. As shown in the fourth column, the purities stay reasonable in the perturbed case.}
		\label{table_scaling}
\end{table}

\begin{figure}[h!]
	\includegraphics[width=0.5\textwidth]{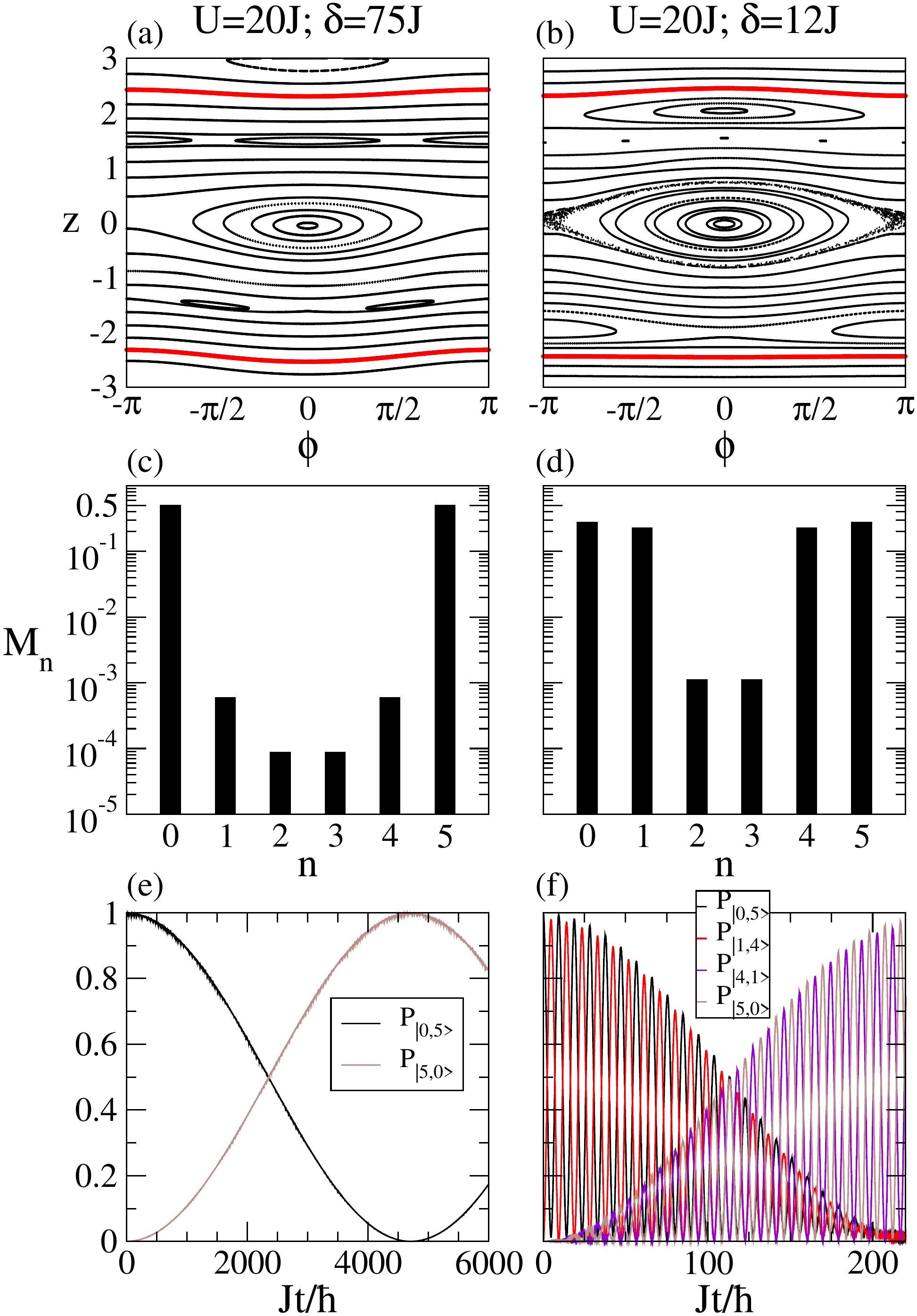}
	\caption{NOON states in a near integrable regime, for $U/J=20$. (a,b) Stroboscopic sections of the phase space. (c,d) Admixtures of quasimodes. (e,f) Time evolution of the detection probabilities. The left column represents the case where a 2:1 resonance situated near in $z_{2:1}=1.5$ couples the quasimodes $\ket{5,0}$ and $\ket{3,2}$ ($\delta/J=75$ and $\hbar\omega/J=120$) while the right column represents the case where the 1:1 resonance located in $z_{1:1}\simeq-2$ couples the quasimodes $\ket{0,5}$ and $\ket{1,4}$ ($\delta/J=12$ and $\hbar\omega/J=80$). For the latter, the two level approximation does not hold and the dynamic takes place on two time scales (see the relation (\ref{double_osci_analytic}) with the panels (d) and (f)). For the 2:1 resonance case, the NOON time is given by $\tau=2.4\times 10^3\hbar/J$, with an excellent purity $p=0.999$, while $\tau_s$ is equal to 112.3$\hbar/J$ in the 1:1 case.
	}
	\label{21_11_res}
\end{figure}

\section{NOON state in the near integrable regime}
As discussed above, the presence of chaos tends to decrease the purity, even though this one remains reasonably good as shown in the figure \ref{proba_mesure_chaos} and in the table \ref{table_scaling}. It is therefore interesting to point out that chaos in the phase space is not necessarily required to decrease the NOON time. This section is devoted to the study of the effect of nonlinear resonances without chaos on the NOON time. The left column of the figure \ref{21_11_res} displays a 2:1 resonance located in $z_{2:1}\simeq1.5$. The transition of $\ket{5,0}$ to the opposite part of the phase space is facilitated by the coupling with the quasimode $\ket{3,2}$. By means of the relation (\ref{external_frequency}), the external frequency that must be applied amounts to $\hbar\omega/J=120$. A procedure similar to the one used in the figure \ref{NOON_time} enables to determine $\delta/J=75$. With a NOON time $\tau=2.4\times 10^3\hbar/J$ (and for $\ce{^87{Rb}}$, $\tau=11$s), the reduction is effective but limited while keeping an excellent purity (see last panel of the left column of the figure \ref{21_11_res}). Indeed, we obtain $p\simeq 0.999$, which is, from this point of view, an improvement in comparison with the mixed regular-chaotic case (see the right column of figure \ref{proba_mesure_chaos}).

It is actually possible to produce a NOON state in the near-integrable regime with the same time scale as in the mixed regular-chaotic case (see panels (c) and (f) of the figure \ref{experimental_simulation}), namely by tuning the driving frequency to an exact resonance in the quantum Floquet spectrum. However, some new complications arise. As an example, a transition between $\ket{0,5}$ and $\ket{5,0}$ can be enhanced by means of a 1:1 resonance situated in $z=2$ (see right column of the figure \ref{21_11_res}). In this case the coupling matrix element with the quasimodes $\ket{1,4}$ and $\ket{4,1}$ is enhanced. In order to have a maximal effect, i.e. the lowest tunneling time $\tau$, we obtain $\delta/J\approx 12$ in that case. For this value, the coupling between $\ket{0,5}$ and $\ket{1,4}$ is so strong that the dynamics must modelled by four-level oscillations (the system is initially in $\ket{0,5}$):
\begin{align*}
&\ket{\phi(t)} 
\simeq\\&\cos\left(\frac{\Omega_s}{2} t\right) 
\left[ \cos\left(\frac{\Omega_f}{2}t\right)\ket{0,5} - i \sin \left(\frac {\Omega_f}{2} t\right) \ket{1,4} \right] \\
&-i \sin\left(\frac{\Omega_s}{2} t\right) 
\left[ \cos\left(\frac{\Omega_f}{2}t\right) \ket{5,0}- i \sin \left(\frac {\Omega_f}{2} t\right) \ket{4,1} \right] \numberthis \label{double_osci_analytic}
\end{align*}
The system presents two different frequency scales given by
\begin{align}
&\hbar\Omega_s = \dfrac{\epsilon_4^+-\epsilon_3^- + \epsilon_2^+-\epsilon_1^-}{2}\\
&\hbar\Omega_f = \dfrac{\epsilon_4^+-\epsilon_2^- + \epsilon_3^+-\epsilon_1^-}{2}
\end{align}
with $\epsilon_1^-<\epsilon_2^+<\epsilon_3^-<\epsilon_4^+$ and where we assume $\epsilon^+_4 - \epsilon^-_3 \simeq \epsilon^+_2 - \epsilon^-_1 $.
 The lower panel of the right column of the figure \ref{21_11_res} displays the probabilities of measuring the different site populations knowing that the system is initially in $\ket{0,5}$. For this particular combination of parameters ($U/J=20$, $\delta/J=12$ and $\hbar\omega/J=80$), the two doublets which mainly contribute to the dynamics present the following energy values:
\begin{align*}
&\epsilon_4^+= 200.435 J\\
&\epsilon_3^-= 200.419 J\\
&\epsilon_2^+=199.767 J\\
&\epsilon_1^-=199.754 J.
\end{align*}
Each of these four levels corresponds to a strong admixture between $1/\sqrt 2 (\ket{0,5}\pm\ket{5,0})$ and $1/\sqrt 2 (\ket{1,4}\pm\ket{4,1})$. They determine the characteristic times of the slow oscillations $\tau_s={\pi}/{(2\Omega_s)}=112.3 \ \hbar/J$ and the fast oscillations $\tau_f ={\pi}/{(2\Omega_f)}= 2.357 \ \hbar/J$. The former time scale determines the time required to produce a NOON state, i.e. the NOON time. For experimental parameters of $\ce{^87{Rb}}$, the NOON time becomes $\tau_s= 0.49$s, which corresponds to the NOON time scale of the mixed regular case (see panel (c,f) of the figure \ref{experimental_simulation}).

\begin{figure}[!h]
	\includegraphics[width=0.4\textwidth]{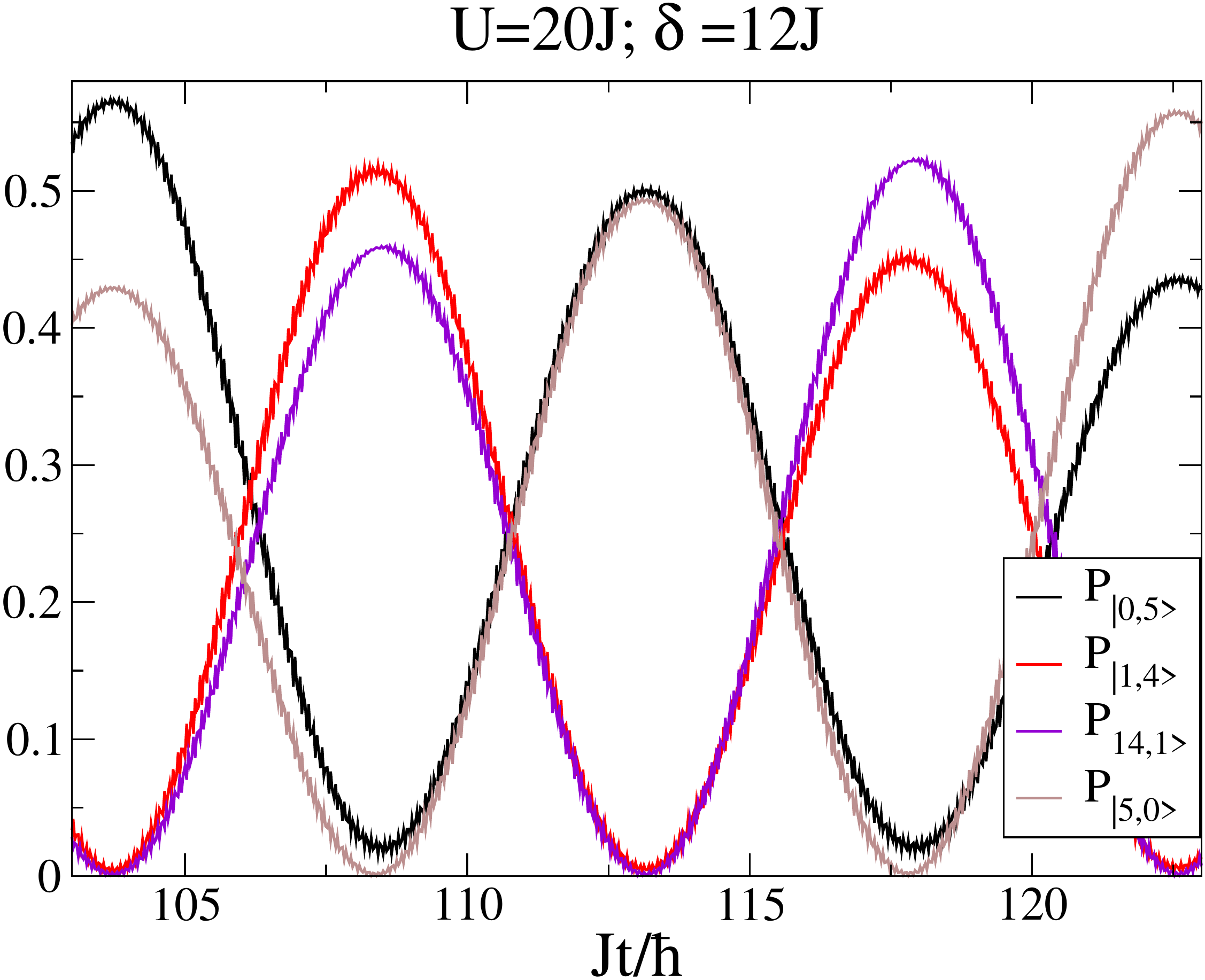}
	\caption{Zoom around $\tau_s$ on the detection probabilities for the 1:1 resonance case (see panel (f) of the figure \ref{21_11_res}). Even though this configuration is not able to produce a perfectly unbiased NOON state ($P_{\ket{0,5}}=P_{\ket{5,0}}=0.5$), a slightly biased NOON state with almost no impurity, i.e. $P_{\ket{0,5}}+P_{\ket{5,0}} \simeq 1$ and $P_{\ket{0,5}}-P_{\ket{5,0}}=8.5\times 10^{-3}$,   can be realized by choosing the measurement time slightly larger than $\tau_s=112.3\hbar/J$, namely $t_m=113.25\hbar/J$. 	
	}
	\label{zoom_11}
\end{figure}

In order to obtain a perfectly balanced NOON state without impurities, i.e. with $P_{\ket{0,5}}=P_{\ket{5,0}}=0.5$, on time $\tau_s$, the frequency ratio $\Omega_f/\Omega_s$ must by divisible by four according to the relation (\ref{double_osci_analytic}). Fine tuning of the amplitude of the driving $\delta$ is a way to achieve this purpose, but this may become very difficult in practice. Alternatively, if $\Omega_f/\Omega_s$ is not exactly a multiple of four (such as $\Omega_f/\Omega_s=47.62$ in our case), it is possible to create a NOON state without impurities ($P_{\ket{0,5}}+P_{\ket{5,0}}=1$), but with a small bias between $P_{\ket{0,5}}$ and $P_{\ket{5,0}}$. By inspecting the zoom around $\tau_s$ shown in the figure \ref{zoom_11}, the measurement time, $t_m$, can be chosen "by hand" in order to meet these criteria. For example, if we choose a $t_m=113.25\hbar/J$ slightly larger than $\tau_s$, we obtain $P_{\ket{0,5}} \approx P_{\ket{5,0}} \approx 0.5$ with $P_{\ket{0,5}}- P_{\ket{5,0}} = 8.5\times 10^{-3}$ and $P_{\ket{4,1}}\approx P_{\ket{1,4}} \approx 10^{-3}$. Knowing that $\Omega_f \approx 0.03 \omega$, the control of the measurement time $t_m$ must be precise on the time scale 10$T$, with $T=2\pi/\omega$, in order to reach a precision on $\tau_f$.
 
\section{Conclusion}
In summary, this paper has presented  numerical evidence that microscopic NOON states can be realized on a realistic time scale with ultracold bosonic atoms trapped in a double well potential. The production of a giant entangled state in a double well potential through collective tunneling displays a major obstacle, namely a very large NOON time exceeding the life time of a condensate. This obstacle can be bypassed by an external perturbation. While the NOON time is prohibitive in the unperturbed case, the addition of a periodic shaking enables to lower the NOON time by several orders of magnitude. The resonant parameters for both the amplitude and the frequency of the driving can be understood and fixed through chaos- and resonance-assisted tunneling. The combination of a central chaotic sea and nonlinear resonances correctly placed in the phase space is able to facilitate the transition through the dynamical barriers. In particular, the paper showed that it is possible  to reach, for five particles, the coherent superposition between $\ket{0,5}$ and $\ket{5,0}$ after a time scale of $\sim$1s, while keeping a relatively good purity of roughly 99 percents. Moreover, the slope of the NOON time as a function of $N_p$ is lower compared to the unperturbed case. This opens perspectives for realizing Schr\"odinger cat states with an increasing number of atoms.
 
An aspect that comes into play in that case and that we did not focus on in this paper is the possibility to realize the NOON superposition not only via tunneling of an atom to the ground mode of the initially empty well, but also via tunneling to one of its excited modes \cite{Garcia-March2011,Garcia-March2012,Gillet2014}. As is detailed in Appendix A, this scenario is not likely to occur for the choice of parameters that we specifically consider in this study. However, for larger atomic populations, such transitions to excited modes will constitute important additional channels through which the NOON tunneling process will take place. Even though a semiclassical treatment of resonance- and chaos-assisted tunneling will become more complicated in that case \cite{Firmbach2019}, one can safely assume that the near-resonant access to transversally excited states in the well will generally give rise to a further enhancement of the NOON tunneling rate and hence to a further reduction of the time needed to produce the NOON superposition.

Another important obstacle that we did not discuss here is to maintain a nearby perfect symmetry between the two sites of the optical lattice on a time scale of seconds. 
A small shift due to, for example, gravity can introduce a perturbation larger than the matrix element coupling $\ket{N_p,0}$ and $\ket{0,N_p}$, which could inhibit the collective tunneling. One way to overcome this problem could be to employ a time crystal for which the symmetry is guaranteed by construction \cite{Sacha2015}. Specifically, on could consider that the atoms are on a ring-shaped trap \cite{Shin2005} in a rotational state, which is generated by a 2:1 resonance via an external driving, i.e., all atoms are prepared within one of the two rotational motion states that are stabilized by the 2:1 resonance of the driving. After the NOON time, the coherent superposition occurs between this initial state and the opposite one on the ring-shaped trap. Further studies could be done in this direction in order to meet the experimental challenge of achieving the perfect symmetry.

\vspace{1cm}

We thank A. B\"acker and R. Ketzmerick for useful discussions and for having welcomed G.V. on a research stay in their Computational Physics group at TU Dresden. Financial support from the Belgian F.R.S.-FNRS (FNRS aspirant grant for G.V.) is gratefully acknowledged.

\appendix
\section{Parameter evaluation}\label{an_parameter_evaluation}
The NOON times presented in the figure \ref{NOON_time} are dimensionless. It is possible to calculate an approximate time in physical units based on some assumptions. The Hamiltonian of a single particle trapped in a homogeneous 1D optical lattice surrounded by an harmonic confinement reads
\begin{equation}
H=-\dfrac{\hbar^2}{2m} \Delta+V_0\dfrac{\hbar^2k^2}{m} (1-\cos(kx)) + \dfrac 1 2 m \omega_\perp^2 (y^2+ z^2).
\end{equation}

If the oscillations inside the well are small, an harmonic approximation can be made and the ground state on site $l$ reads $\varphi_l(\vec r)=\varphi(\vec r-l\frac{2\pi}{k}\vec e_x)$, where
\begin{equation}
\varphi(\vec r) \simeq \dfrac{1}{\sqrt{\sqrt \pi \sigma_\parallel}} \exp\left(-\frac{x^2}{2\sigma_\parallel^2}\right)  \dfrac{1}{\sqrt \pi \sigma_\perp} \exp\left(-\dfrac{y^2+z^2}{2\sigma_\perp^2}\right),
\end{equation}
with $\sigma_{\parallel} =1/(kV_0^{1/4})$ and $\sigma_\perp=\sqrt {\hbar/(m\omega_\perp)}$. Still in the harmonic approximation, the longitudinal trap frequency takes the form $\omega_{\parallel}=\sqrt{V_0}\hbar k^2/m$. The evaluation of the on-site interaction is given by \cite{Bloch2008}
\begin{equation}
U= g \int_{\mathbb R^3} |\varphi (\vec r)|^4 d^3r \simeq 2\hbar \omega_\perp k a_s \sqrt{\dfrac{\sqrt V_0}{2\pi}},
\label{U_evaluation}
\end{equation}
with $a_s$ the s-wave scattering length, $g=4\pi \hbar^2a_s/m$ 
and $\omega_\perp$ the transverse trap frequency. A semiclassical evaluation (by means of WKB theory) of the transmission coefficient through a potential barrier for a particle in its ground state gives rise to the hopping parameter \cite{Garg2000,Mouchet2006_phd}
\begin{align}
 J&=\dfrac{\hbar\omega_{\parallel}}{\sqrt{e\pi}} \exp\left(- \dfrac 2 \hbar \int_a^{\pi/k} \sqrt{2m(V(x)-E_0)} \ dx \right)\\
 &=\sqrt{\dfrac{V_0}{e\pi}} \dfrac{\hbar^2k^2}{m}\exp\left( -2\sqrt{2V_0} \int_{\arccos \eta}^{\pi} \sqrt{\eta-\cos\phi} \ d\phi \right),
 \label{J_sc}
\end{align}
with $\eta=\exp\left(-1/(4\sqrt{V_0})\right)-1/(4\sqrt V_0)$, $V(x)= V_0 \hbar^2k^2/m (1-\cos(kx))$ and $E_0=V(a)=\sqrt{V_0}(1-\eta)\hbar \omega_\parallel$ the ground state energy of the one dimensional system. By assuming $\omega_\perp$ = $\omega_{\parallel}$, $V_0$ can be determined by solving 
\begin{equation}\label{U_J}
\dfrac U J \simeq  \sqrt{2e\sqrt {V_0} } k a_s \exp\left(2\sqrt{2V_0} \int_{\arccos\eta}^{\pi}\sqrt{\eta-\cos \phi} \ d\phi\right).
\end{equation}
The energy scale is then obtained by means of the relations (\ref{U_evaluation}) or (\ref{J_sc}). 

The wavelength of the laser is assumed to be equal to $\lambda$=1064 nm with $k=2\pi/d=2\pi/(\lambda/2) = 2k_\varphi$ ($d$ is the distance between two wells and $k_\varphi$ the frequency of the laser photons). The s-wave scattering length and the mass of $\ce{^{87}Rb}$ are $a_s=5.313$nm and $m=$1.443$\times 10^{-25}$kg respectively. The solution of (\ref{U_J}) for $U/J=20$ gives $V_0=0.8586$. That leads to $\omega_{\parallel}=\omega_{\perp}=9.4\times10^4$s$^{-1}$ and $\hbar/J=4.4\times 10^{-3}$s. 
A harmonic approximation of the band energies takes the form $E_n\simeq \sqrt{V_0}(n+1/2)\hbar^2k^2/m$ with $n$ the band index. Knowing that the strength of the potential is $2V_0\simeq1.7$, the wells have the lowest energy band near $E_0\simeq0.46\hbar^2k^2/m$. 

The role of the nonlinear resonances and chaos is striking concerning the diminution of the NOON time. We obtain $\tau(\delta/J=0)\simeq 2.6 \times 10^3$s for the unperturbed system whose phase space is displayed in the panel (e) of the figure \ref{experimental_simulation}, $\tau(\delta/J=19.5)\simeq 0.84$s for the 1:4 resonance with the chaotic layer (panel (c) and (f) of the figure \ref{experimental_simulation}), $\tau(\delta/J=75)\simeq11$s for the case with the 2:1 resonance (left column of the figure \ref{21_11_res}) and $\tau_s(\delta/J=12)\simeq 0.49$s for the case with the 1:1 resonance (right column of the figure \ref{21_11_res}).

A condition for the validity of our semiclassical approach is that the particle removal energies of the sites (which are identical to their local chemical potentials when their populations are large) have to be sufficiently low to avoid a transition to the first excited band situated at $E_1\simeq 1.39 \hbar^2k^2/m $. The particle removal energy of site $i$ ($\mu_i$ with $i=1,2$) is largely dominated by the on-site interaction, giving rise to the relation $\mu_i \simeq U  (n_i-1)$  (for $n_i\geqslant 1$). For the quasimode $\ket{0,5}$ the energy cost to remove one particle reads $\mu_2^{(0,5)}\simeq 0.18 \hbar^2k^2/m $. As $E_1-E_0\simeq 0.93 \hbar^2k^2/m$, the transition of one particle to the first excited band is avoided. Note that virtual transitions to such an excited state can nevertheless play a nonnegligible role in the case that the population on the site under consideration is increased and the particle removal energy from the occupied site therefore becomes closer to the transition to the excited state. The main effect of such virtual transitions will be to open up a secondary tunneling channel for the NOON doublet under consideration. From the theory of resonance-assisted tunneling in higher dimensional systems \cite{Firmbach2019} we should expect that the presence of this secondary channel will generally give rise to an additional enhancement of the tunneling rate, and hence to a further reduction of the time needed to produce the NOON superposition, as compared to the study undertaken in this paper. The semiclassical and numerical calculations will then become more involved \cite{Firmbach2019} even though the end result, concerning the reduction of the NOON time, is expected to be qualitatively very similar to the present study.


\section{Semi-classical evaluation of the NOON time} \label{sc_evaluation}
In the framework of resonance-assisted tunneling, the dynamics near a $r$:$s$ resonance can be approximated by a pendulum, whose the classical Hamiltonian is written \cite{Schlagheck2011}
\begin{equation}
   H_{\text{res}}^{(r:s)} (I, \theta) = \dfrac{(I-I_{r:s})^2}{2m_{r:s}} + 2V_{r:s}\left(\dfrac I {I_{r:s}}\right)^{r/2} \cos(r\theta+\phi_{r:s}).
\end{equation}
$V_{r:s}$ is the potential amplitude of the pendulum, $I_{r:s}$ its action, $m_{r:s}$ its effective mass and $\phi_{r:s}$ is an arbitrary phase. The coupling matrix elements between quasimodes are introduced by means of a semi-classical quantization of $H_{\text{res}}^{(r:s)}$.
\begin{equation}
\langle n-r | \hat H_{\text{res}}^{(r:s)} \ket n = V_{r:s} \e ^{-i\phi_{r:s}} \left(\dfrac{\hbar}{I_{r:s}}\right)^{r/2} \sqrt{\dfrac{n!}{(n-r)!}}
\end{equation}

 In the integrable case, the two symmetric parts are connected by a prominent central island (see panel (e) of figure \ref{experimental_simulation}), which produces the unperturbed splittings, $\Delta \epsilon_n^{(0)} =|\epsilon_n^{(0)-} - \epsilon_n^{(0)+}| $. The question is how the $r$:$s$ resonance influences the tunneling between the two symmetry-related parts of the phase space. For $N_p=5$, there are 3 quantum eigenstates ($n=0,1,2$) within each symmetry subspace, knowing that the NOON state refers to $n=2$. The quantum perturbation theory in the near-integrable regime gives rise to \cite{Schlagheck2011}
 \begin{align}
 &\Delta \epsilon_n \simeq \Delta\epsilon_n^{(0)} + |\mathcal A_{n,n-r} |^2 \Delta \epsilon_{n-r}^{(0)}, \label{sc_DE} \\
 &\mathcal A_{n,n-r} = \dfrac{\bra{n-r} \hat H_{\text{res}}^{(r:s)} \ket{n} }{\epsilon_n^{(0)}-\epsilon_{n-r}^{(0)}-s\hbar\omega}. \label{Ars}
 \end{align}
  
 The $r$:$s$ = 2:1 case (see the left column of the figure \ref{21_11_res}) fulfills the assumption. The amplitude $V_{2:1}=0.0239$ and the action $I_{2:1}=1.5014$  are directly extracted from the phase space \cite{Eltschka2005}. The unperturbed energies from the antisymmetric block are used in the denominator of (\ref{Ars}) (it could have been the symmetric block). The semiclassical evaluation of the NOON time reads
 \begin{equation}
 \tau = \dfrac{\pi\hbar}{2 \Delta\epsilon_2} \simeq 3.8\times10^3 \ \hbar/J,
 \end{equation}
 compared to the exact NOON time $\tau=2.4\times 10^3 \ \hbar/J$.
 
In a mixed regular-chaotic system, the chaotic sea, which can be seen as the phase-space regions where the quasimodes are strongly mixed, must be taken into account. In this context, the chaotic part of the spectrum can be modeled by the Gaussian orthogonal ensemble (GOE). The resulting effective Hamiltonian matrix describing resonance- and chaos-assisted tunneling can be cast in the form
\begin{equation}
\begin{pmatrix}
\begin{tikzpicture}
\matrix (m)[
matrix of math nodes,
nodes in empty cells,
nodes={text width={width(998)}, minimum width=9mm, align=center, minimum height=6mm}
] {
	E_n   &  V_{\rm eff}  & 0 &\cdots & 0 & \Delta_n \\
	V_{\rm eff} &    &  &   &   & 0  \\
	       0    &    &  \text{chaotic sea} &   &     & \vdots     \\
	     \vdots &    & \rm (GOE) &   &   & 0 \\
	       0    &    &  &   &   & V_{\rm eff}\\
	\Delta_n    &   0 &\cdots & 0 & V_{\rm eff}   & E_n    	\\
} ;
\draw (m-2-2.north west) rectangle (m-5-5.south east) ;
\end{tikzpicture}
\end{pmatrix},
\end{equation}
with $E_n$ the energy in absence of hopping, $V_{\text{eff}}$ the effective coupling matrix element between the NOON state and the chaotic sea, and $\Delta_n=\Delta \epsilon^{(0)}_n/2$. The square symbolizes a full Hermitian matrix displaying a discrete symmetry with respect to the counter-diagonal, due to the presence of the parity. This matrix describes the chaotic part of the phase space and can be modeled by GOE. In this framework, one can show that the level splittings are statistically distributed according to a Cauchy distribution \cite{Leyvraz1996}. By defining a suitable cutoff, one can show that the logarithmic mean $\langle \Delta\epsilon \rangle_g =\exp(\langle \ln\Delta\epsilon \rangle)$ is given by \cite{Schlagheck2011}
\begin{equation}
\langle \Delta\epsilon \rangle_g = \dfrac{2\pi V_{\text{eff}}^2}{\hbar \omega}.
\end{equation}

An additional ingredient is needed in order to accurately describe the tunneling process in the mixed regular-chaotic case, namely the presence of partial barriers \cite{Schlagheck2011,Bohigas1993,BohigasTom1993}, which tile the chaotic sea into subregions that are not well connected. For example, the homoclinic tangle of the 1:3 resonance in the panel (f) of the figure \ref{experimental_simulation} defines barriers in the chaotic sea that have to be taken into account to evaluate the splitting. 

In this particular case, tunneling mainly takes place via a two-resonance process, involving the resonances $r_1$:$s_1$=$1$:$4$ ($I_{1:4}=2.004$ and $V_{1:4}=0.06857$) and $r_2$:$s_2$=$1$:$3$ ($I_{1:3}=1.511$ and $V_{1:3}=0.3017$). The effective coupling can be evaluated by means of the dominant contribution,
\begin{align}
V_\text{eff}^2 = \dfrac{\left|V_{r_1:s_1}^{( n-r_1)}\right|^2}{\left|\epsilon^{(0)}_n-\epsilon^{(0)}_{n-r_1}-s_1\hbar\omega\right|^2}   \left|V_{r_2:s_2}^{(n-r_1 -r_2)}\right|^2,
\end{align}
with $V_{r:s}^{(\nu)} = \langle \nu| \hat H_\text{res}^{(r:s)}| \nu+r \rangle$. For $N_p=5$, the NOON state refers to $n=2$. Then, the NOON time is semiclassically evaluated as 
\begin{equation}
\tau = \dfrac{\pi\hbar}{2\langle \Delta \epsilon\rangle_g} \simeq 1.4\times 10^2 \ \hbar/J,
\end{equation} 
compared to the exact time $\tau=1.9\times 10^2 \ \hbar/J$.

\bibliographystyle{apsrev4-1} 

\bibliography{biblio} 


\end{document}